\documentclass[review]{elsarticle}
\usepackage{lineno,hyperref}

\usepackage{xcolor}
\usepackage{ragged2e}
\usepackage{graphicx}
\usepackage{pbox}
\usepackage{amsmath}
\usepackage{tabularx}
\usepackage{subfig}
\usepackage{graphicx}

\usepackage{amsmath, amsfonts, bm, color, ulem}
\usepackage{units}
\usepackage{breqn}

\usepackage{comment}
\usepackage[english]{babel}

\makeatletter
\def\ps@pprintTitle{%
  \let\@oddhead\@empty
  \let\@evenhead\@empty
  \def\@oddfoot{\reset@font\hfil\thepage\hfil}
  \let\@evenfoot\@oddfoot
}
\makeatother

\begin{document}
\nolinenumbers
\begin{frontmatter}

\title{Energy implications of a load depending on geometrical configurations in an oscillator}


%\author{Elena, Jimena, Miguel, Almudena, Laura, Andres, Daniel, Javier, Rubén.\fnref{myfootnote}}
%\address{BioCoRe, Biological Cooperative Research S Coop.}

\author[1]{Elena Campillo Abarca}
%\ead{elena.campillo@biocore.es}

\author[1]{Jimena de Hita Fernández,}
\author[1]{Almudena Martínez Cedillo,}
\author[1]{Miguel León Pérez}
\author[1]{Laura Morón Conde}
\author[1]{Andrei Sipos}
\author[1]{Daniel Heredia Doval}
\author[1]{Javier Domingo Serrano}
\author[1]{Rubén González Martínez}
%\author[1]{Rubén González Martínez}
%\ead{ruben@biocore.es}

\address[1]{BioCoRe, Biological Cooperative Research S.Coop, Madrid, Spain}

%\ead[url]{https://biocore.es/}



\begin{abstract}
    
This paper studies, for a specific oscillatory system composed by a pendulum connected to a seesaw, how the geometry of the different mechanisms of energy introduction conditions the resulting movement, to achieve both a greater amplitude of oscillation due to a change of velocity and an acceleration in its movement. The different configurations that give rise to the acceleration of motion are therefore analyzed. The study is carried out from a kinematic point of view, theoretically simulating an energy increase in the system and analyzing its response in terms of angular velocity and of modification of apparent weight. Subsequently, the force diagram necessary to generate the accelerated motion is analyzed. The magnitude of the external force to be applied and its dependence on the direction and angular instant in which it are exerted is evaluated. It is observed how for some specific configurations this magnitude is negative, implying that the system is capable of accelerating when subjected to a brake or load on it.

\end{abstract}
\end{frontmatter}
\nolinenumbers

\section{Introduction}

An oscillatory system can illustrate a variety of physical behaviors with the advantage that the starting point is more accessible \cite{FundamentalsofPhysics}. The pendulum system is not a recent discovery, it has been extensively analyzed since the study of pendulum isochronism by Galileo \cite{cespedesprimer}  who gave account of the dependence relation between the period and the amplitude of a pendulum's oscillation \cite{galileo}. The present work is part of a series of investigations that provide a clear construction, through the study of an oscillator, of fundamental conceptual objects that can be extrapolated to more complex rotors.

The starting point is a seesaw system with a simple pendulum at one end. First it will be explained how the application of forces to this pendulum can adopt different geometrical configurations. Subsequently, the complete system is considered, showing that the limit given by the oscillation amplitude will mark the apparent weight variations. 

The study then proceeds to analyze the response of the system's oscillatory movement when exerting on it modifications of the environment that provoke the acceleration of the rotation \cite{galileo,modinoemmy}. Two acceleration mechanisms are taken into account: the direct supply of energy and the application of an external force. It is shown that, for some specific geometrical configurations, the oscillatory motion of the system accelerates even if a brake is applied.

\section{Methods}

To begin with, two possible ways of applying a thrust force on a pendulum to increase the amplitude of oscillation will be presented. On the one hand, it is possible to apply the thrust force in a tangential direction with regard to motion. On the other hand, it is possible to obtain the same resultant by applying this force in a normal direction. This second option implies that it is possible to produce such an increase in amplitude through a displacement of the center of mass. 

Next, the response of the oscillator, consisting of a seesaw with a pendulum at one end (Fig.\ref{fig1}), is simulated in different dynamic rotational configurations. The investigation is therefore approached through two procedures. 

First, the response to the direct introduction of energy into the oscillator will be studied. The relationship between the angular velocity acquired by introducing energy into the system and the dynamic balance necessary to maintain the accelerated rotation of the oscillator will be analyzed. In both cases its expression in terms of apparent weight will be considered.
On the other hand, the different geometries of direction and point of application of the force necessary for the acceleration of the system, as well as the possible geometric results of this acceleration, will also be analyzed. 

To perform the calculations, the rotation of the pendulum is fixed on the plane perpendicular to the ground and the zero value of potential is set as the value corresponding to the position of the oscillator $\theta=0^\circ$. Likewise, due to its lack of quantitative and qualitative impact, the dispersion effects due air and rotation axis friction are disregarded.
\section{Result and discussion}

\subsection{System description}

\begin{figure}[htb!]
\centering
\includegraphics[width=90mm]{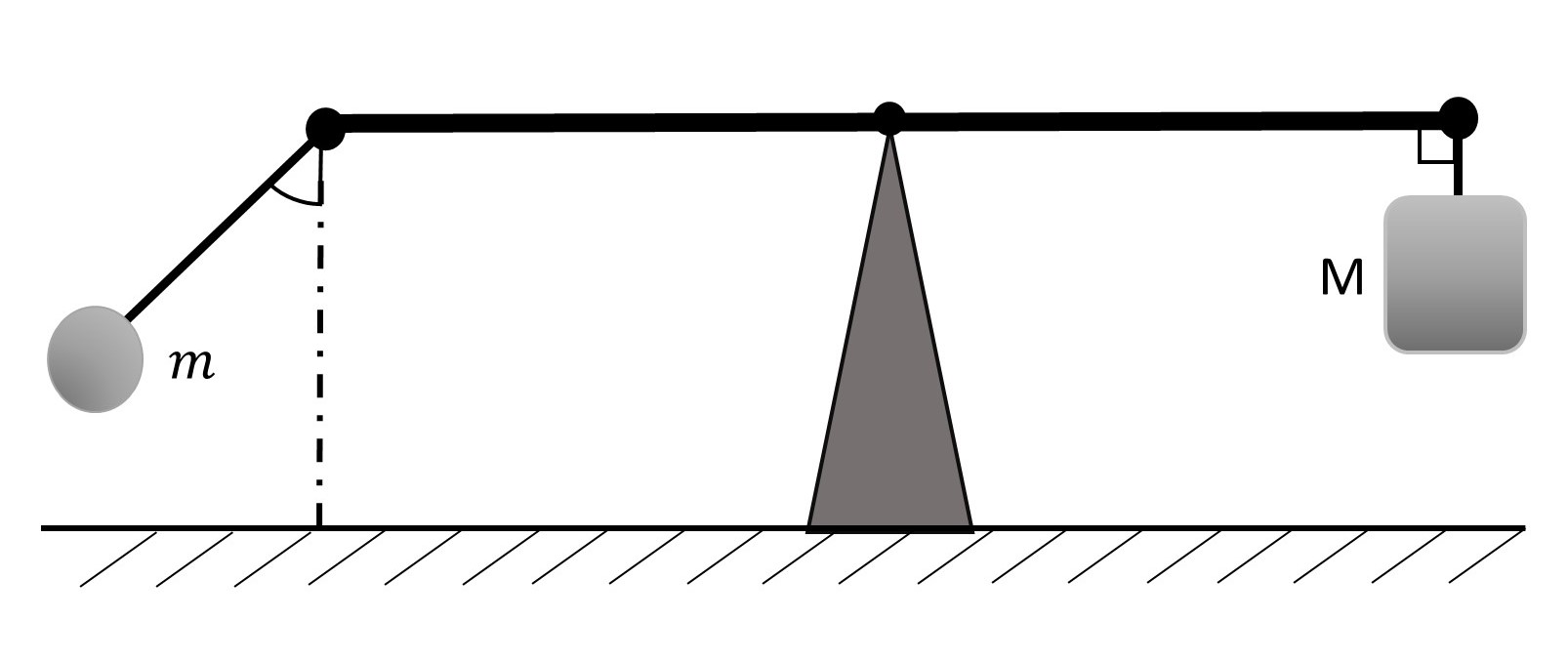}
\caption{Description of the oscillating pendulum system. Where $m$ corresponds to the mass of the oscillating pendulum and $M$ to the variable mass.}
\label{fig1}
\end{figure}

A system that combines a seesaw and a simple pendulum of mass $m$, with a variable mass $M$ to keep the system in dynamic equilibrium at all times (Fig.\ref{fig1}), is supposed.

In addition, the position at which the pendulum is in stable equilibrium, i.e., where its kinetic energy due to motion is maximum ($\theta=0^\circ$), is set as zero potential. The forces involved in the motion of the oscillator are the weight of the oscillating mass $m$, the centrifugal force it acquires due to its inertia \cite{libropenduloresumen} and the weight of the variable mass $M$. In the present analysis air and rotational axis friction are disregarded. As a consequence, the oscillator will have a uniform angular velocity which will change only if some force is introduced.

\subsection{Introduction of forces in the system} 

\begin{figure}[htb!]
\centering
\includegraphics[width=60mm]{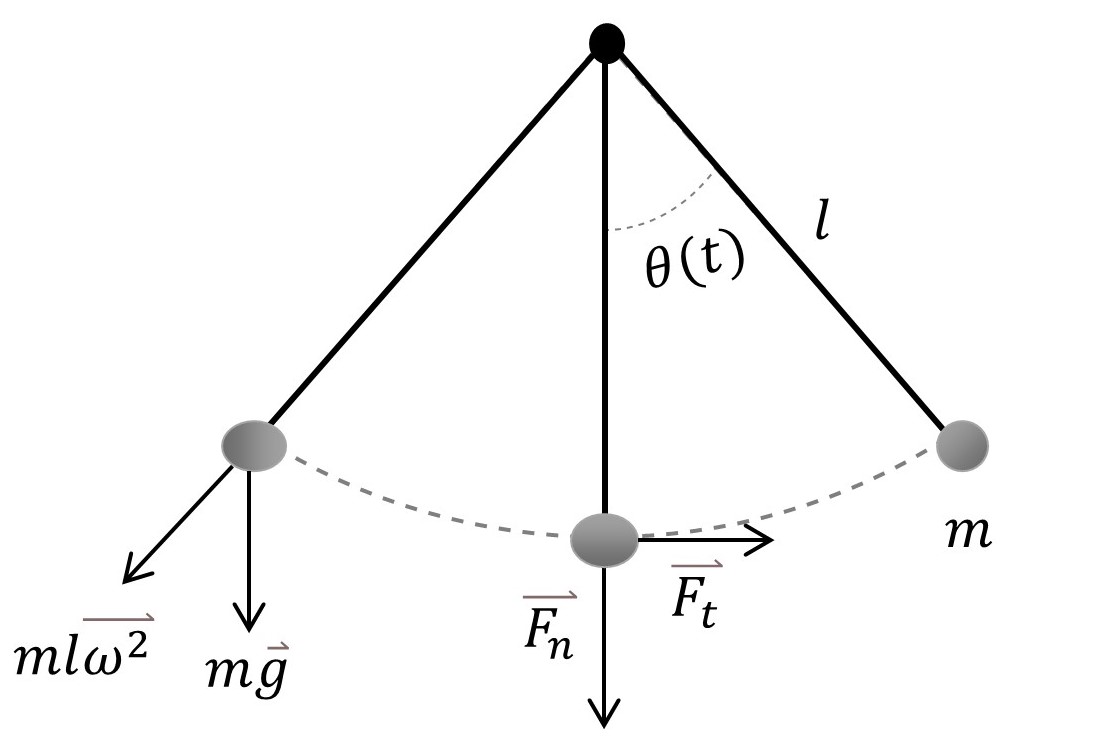}
\caption{Description of the oscillatory system with motion given by $\theta(t)$. Where $m$ is the mass of pendulum and $l$ is the pendulum length, $ml\vec{\omega^2}$ corresponds to the centrifugal force and $m\vec{g}$ to the weight. Visualization of the normal and tangential forces $\vec{F_{n}}$ and $\vec{F_{t}}$.}
\label{fig2}
\end{figure}

In the first stage of the analysis, only the part of the oscillator comprising the simple pendulum will be studied (Fig.\ref{fig2}). The forces involved in its motion are the weight of the mass $m$ rotating at $\theta(t)$ and the centrifugal force it acquires due to its inertia \cite{libropenduloresumen}. Along the movement, the weight remains in normal direction with regard to the ground while of the centrifugal force always has a radial direction \cite{FundamentalsofPhysics}. The work that these forces will have to do in order to introduce a given energy increase in the system depending on the point and direction in which they are applied will now be analyzed. The different dynamic configurations to be considered will be the following: on the one hand, the application of a force in a tangential direction (Eq.\ref{eq:1}) with regard to motion by means of an external  thrust; and, on the other, the appllication of a force in a normal direction (Fig.\ref{fig2}) with regard to motion  that modifies the position of its center of mass. 

\hspace{5mm}

\break

\begin{equation}
\vec{F_{t}}=m\vec{a_{t}}=-mgsen\theta(t)\widehat{t}
\label{eq:1}
\end{equation}

\begin{equation}
\vec{F_{n}}=m\vec{a_{n}}=(mgcos\theta (t)+mlw^{2})\widehat{n}
\label{eq:2}
\end{equation}

\hspace{5mm}

\begin{figure}[htb!]
\centering
\includegraphics[width=70mm]{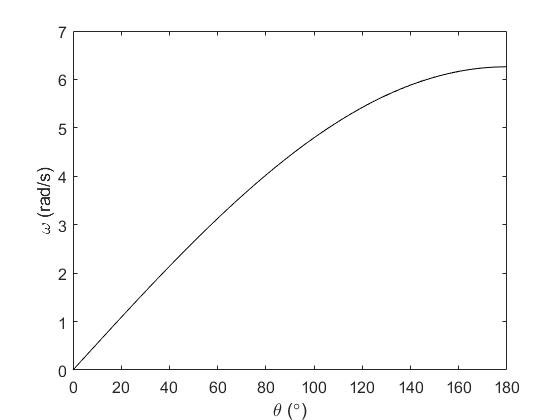}
\caption{Dependence between the angular velocity $\omega$ and the angular amplitude $\theta_0$ from which the pendulum motion starts.}
\label{fig3}
\end{figure}

To obtain the angular velocity of the system, $\omega$, a kinematic analysis of the motion is required: the energy balance at two points of the path, those corresponding to the maximum and minimum potential is evaluated (Ec.\ref{eq:3}) \cite{FundamentalsofPhysics,tipler2021fisica}. There is a quadratic dependence between the angular velocity of the oscillation and the point from which its motion starts, represented by $\theta_{0}$ (Ec.\ref{eq:4}) or, what is the same, between the angular velocity and the angular amplitude of the motion (Fig.\ref{fig3}).

\begin{gather}
\left.\begin{matrix}
E_0=m l g (1-cos{\theta}_{0})\\
E_1=\frac{1}{2}ml^{2}\omega^{2} \\
\end{matrix}\right\}
  E_0=E_1
\label{eq:3}
 \end{gather}
 
\begin{equation}
{\omega^2}= \frac{2g}{l}(1-cos{\theta}_{0}) 
\label{eq:4}
\end{equation}

\begin{figure}[htb!]
\centering
\includegraphics[width=50mm]{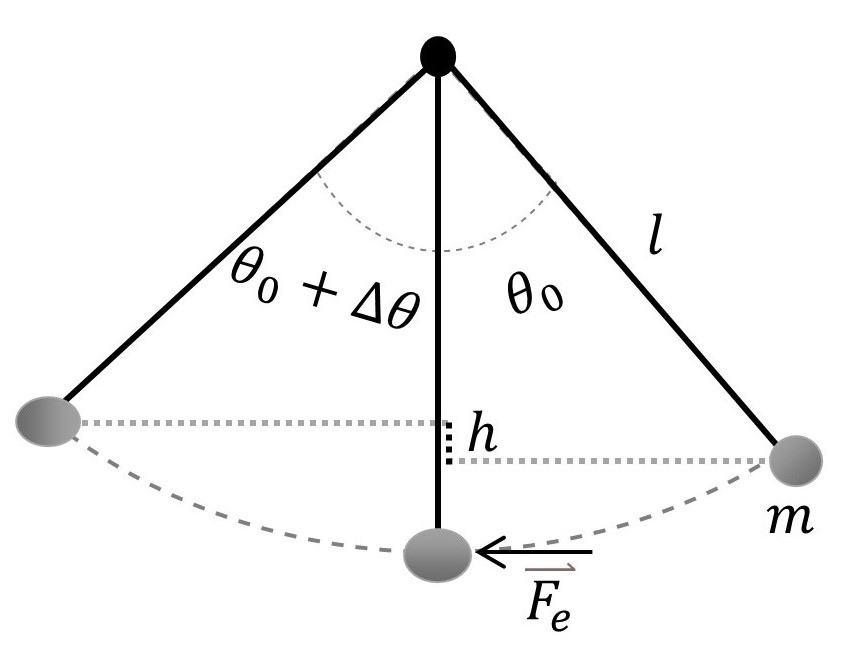}
\caption{Representation of the transformation in which an increment $h$ is reached with an increase in the amplitude of oscillation $\Delta\theta$, due to the application of a tangential force with regard to the motion $\vec{{F}_{e}}$ on the mass $m$. }
\label{fig4}
\end{figure}

In Fig.\ref{fig4}  the scheme of the transformation when applying a tangential force with regard to motion when the oscillator is at the point of its maximum kinetic energy is showed. In this first scenario, in order to increase the amplitude of the oscillation by an angle $\Delta\theta$ the force must overcome the tangential component of the pendulum's weight (Ec.\ref{eq:5}).

\begin{equation}
\vec{F_{e}}=(mgsin(\theta _{0})+mgsin(\theta _{0}+\Delta \theta ) )\widehat{t}
\label{eq:5}
\end{equation}

The potential energy variation between the height achieved after the angular increment $\Delta\theta$ (Ec.\ref{eq:6}) and the height the pendulum would achieve if it naturally oscillated from $\theta_0$ \cite{empuje} would be equal to the work done by applying a  thrust force tangentialto the motion (Ec.\ref{eq:7}). 

\begin{equation}
h=h_{1}-h_{0}= l(cos(\theta_{0})-cos(\theta_{0}+\Delta\theta))
\label{eq:6}
\end{equation}

\begin{equation}
    W_t (\theta _{0},\Delta \theta)= mgh(\theta _{0},\Delta \theta ) = mgl (cos(\theta _{0})-cos(\theta _{0}+\Delta \theta ))
     \label{eq:7} 
\end{equation}

\begin{figure}[htb!] 
\centering
\includegraphics[width=50mm]{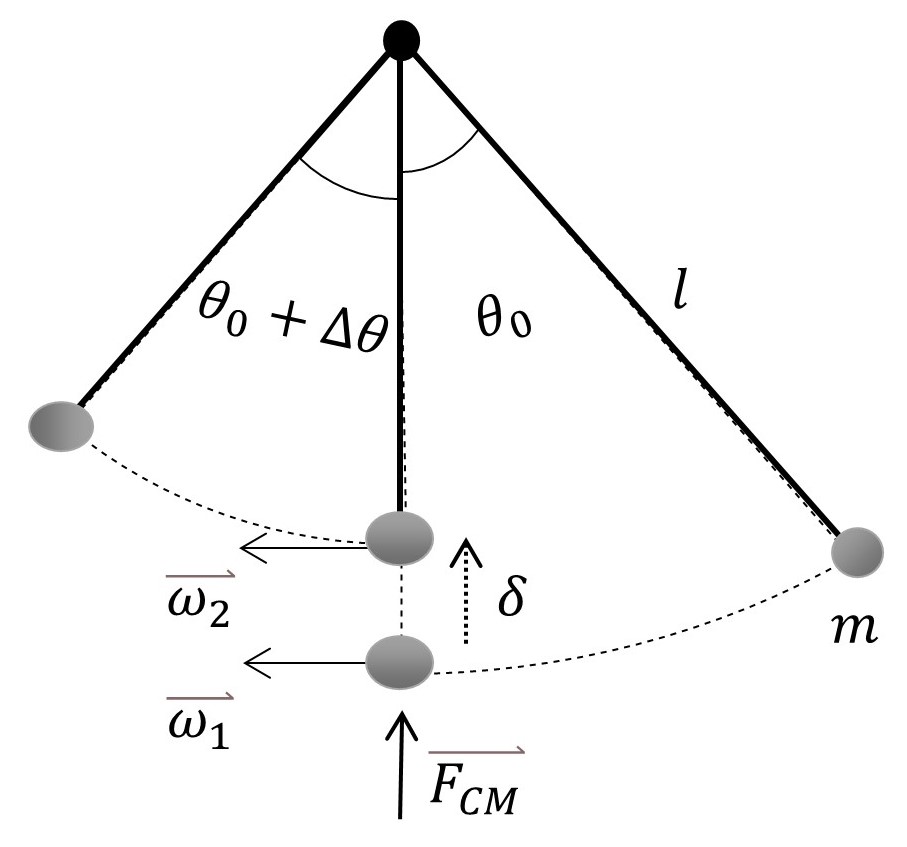}
\caption{Displacement  $\delta$ of the center of mass by the action of an external force $\vec{F}_{CM}$ normal to the pendulum's motion. It is visualized how the translation implies an increase in the angular velocity $\omega$ with a consequent increase in the amplitude of oscillation $\Delta\theta$. }
\label{fig5}
\end{figure}

As mentioned above, a second way to increase the amplitude of the pendulum's movement would be applying a force in a normal direction with regard to motion, displacing the center of mass of the system \cite{columpio,columpio2,columpio3}. Again, the force would be applied at the point of minimum potential energy i.e., at $\theta=0^{\circ}$ (Fig.\ref{fig5}).

Like energy, the magnitude of angular momentum $\vec{L}$ is conserved throughout movement (Eq.\ref{eq:8}). Considering this symmetry at the point where the displacement takes place, the increase in angular velocity (Eq.\ref{eq:9}) that results from such a transformation is obtained  \cite{columpio2,Tea1968PumpingOA}. 
\hspace{-1mm}

\begin{gather}
m{l^{2}}{\omega_{1}}=m(l-\delta )^2\omega_{2}
\label{eq:8}
\end{gather}

\begin{gather}
 \omega_{2}=\frac{l^2}{(l-\delta )^2}\omega_{1} 
 \label{eq:9}
 \end{gather}
 
The displacement of the center of mass of the system, $\delta$, implies an increase in its angular velocity from $\omega_1$ to $\omega_2$ and, consequently, an increase in the oscillation amplitude, $\Delta\theta$. The relationship between the two values, $\delta$ and $\Delta\theta$ \cite{columpio,Tea1968PumpingOA}, is calculated by taking advantage of the other symmetry of the system and the conservation of energy (Ec.\ref{eq:10}).

\begin{equation}
\delta =l-l{\left( \frac{sin\left(\frac{\theta _{0}}{2}\right)}{sin\left(\frac{\theta _{0}+\Delta\theta}{2}\right)}\right) }^{2/3}
\label{eq:10}
\end{equation}

\break

The external force that will be necessary to carry out this transformation (Ec.\ref{eq:11}), and the work to be done in the same direction of displacement of the center of mass overcoming both the inertia component and weight (Ec.\ref{eq:12}) are thus calculated. 

\begin{equation}
     \vec{F}_{CM}=( m(l-\delta)w^{2}+mgcos\theta) \widehat{n}
     \label{eq:11}
\end{equation}
 
%\begin{equation}
\begin{multline}
    W_{n} (\theta _{0},\Delta \theta)= \int_{l}^{l-\delta} -F_{CM} \cdot dx = \int_{l-\delta}^{l} (mw^{2}x+mg)  \cdot dx =   \\  
     = m \omega^2
    (\theta_0) (l  \delta (\theta _{0},\Delta \theta) - \frac{\delta^2}{2}(\theta _{0},\Delta \theta))+mg\delta (\theta _{0},\Delta \theta)
    \hspace*{2cm}
    \label{eq:12}
\end{multline}
%\end{equation}

\subsection{Direct energy input into the system}

Once the geometrical determination of the forces and energy increments in a simple pendulum are known, the direct introduction of energy in the case of the oscillating system can be analyzed. Two different options will be considered: first, a constant increase of angular velocity; then, the introduction of a constant angular acceleration.

\subsubsection{Angular velocity regimes}

The first scenario to be considered is the introduction of an energy increment in the system's motion at its potential energy maximum. The natural oscillation travels through $360^\circ$ with a given angular velocity $\omega$ (Eq.\ref{eq:4}) which depends on the gravity $g$, the pendulum length $l$ and the initial angle $\theta_{0}$. The relation between the change in the pendulum's angular velocity and the magnitude of the energy increment that has been introduced can be obtained from the movement's energy balance (Eq.\ref{eq:13}). That is, by analyzing the kinematic scheme of the system at the points of maximum and minimum potential it is observed that the external energy input will result in a change of the angular velocity (Eq.\ref{eq:14}).

\begin{equation}
\left.\begin{matrix}
E_0= 2 m l g +\Delta E \\
E_1=\frac{1}{2}ml^2{\omega_f}^{2} \\
\end{matrix}\right\}
  E_0=E_1
\label{eq:13}
\end{equation}
 
\begin{equation}
\omega_{f}^{2}=\frac{4g}{l}  +\frac{2}{ml^2} \Delta E =\omega_0^2+\frac{2}{ml^2} \Delta E
\label{eq:14}
\end{equation}

\begin{figure}[h] 
\centering
\includegraphics[width=70mm]{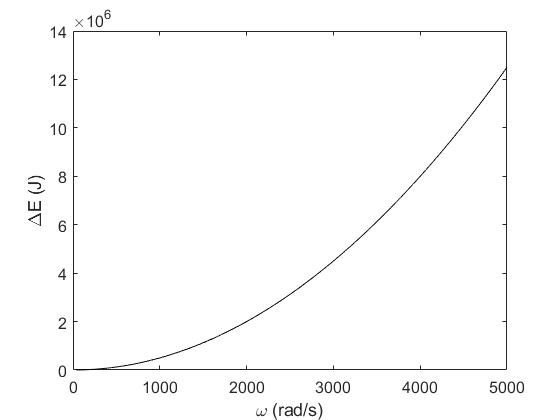}
\caption{ Ratio between the angular velocity $\omega$ of rotation of the pendulum and the energy increment $\Delta E$ introduced into the system. }
\label{fig6}
\end{figure}

As can be seen in Eq.\ref{eq:14} and can be visualized in Fig.\ref{fig6}, there is a quadratic relationship between the magnitude of the energy increment introduced and the resulting angular velocity. Two categories of response can therefore be established depending on the range of angular velocity with which the oscillator rotates. In the environment close to the natural angular velocity the quadratic component will govern the response but, at high angular velocities, the behavior becomes linear.

\begin{figure}[htb!]
\centering
\includegraphics[width=90mm]{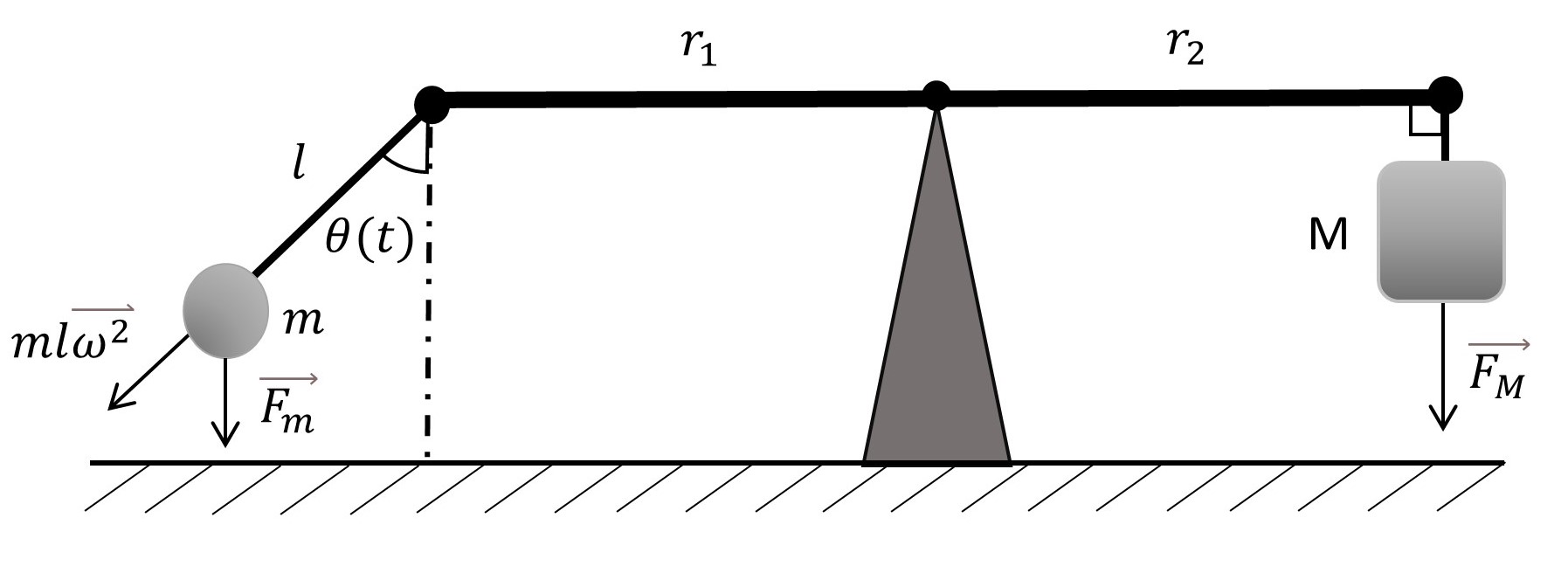}
\caption{Seesaw system. Where $m$ corresponds to the mass of the pendulum swinging due to $\theta(t)$ with length $l$ on one arm $r_1$ with weight $\vec{F_m}$ and inertia $ml\vec{w^2}$. On the other side is the arm $r_2$ in equilibrium depending on its mass $M$ variable, with weight $\vec{F_M}$.}
\label{fig7}
\end{figure}

The fact that the system consists of a pendulum connected to a seesaw (Fig.\ref{fig7}), leads to a closer look at the kinetic phenomenon that is being described. As the seesaw remains in equilibrium condition, the mass $M$, with only vertical motion, must adapt to the variations of apparent weight undergone by the oscillator along its path (Eq.\ref{eq:15}). That is to say, the torque $\vec{\tau}$ experienced by both ends of the system must be compensated, obtaining the expression for the variation of the apparent weight in the system (Eq.\ref{eq:16}). Thus, the increase of the angular velocity of the system due to the introduction of energy into it will modify the apparent weight along the oscillation.

\begin{equation}
r_{1}(ml\omega^{2}cos\theta+mg)=r_{2}Mg
\label{eq:15}
\end{equation}

\begin{equation}
M(t)=\frac{r_{1}}{r_{2}} \left( \frac{l\omega_{f}^{2}}{g}cos\theta(t) +1\right)m
\label{eq:16}
\end{equation}

\begin{figure}[htb!]
\centering
\includegraphics[width=75mm]{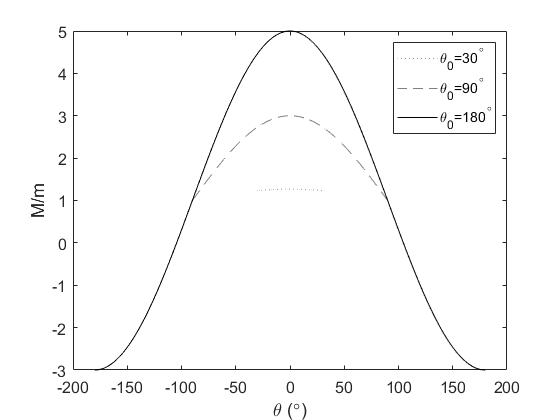}
\caption{Variation of apparent weight along the motion for oscillations of amplitudes of $\theta$ = $60^\circ$, $180^\circ$ and $360^\circ$. That is, for initial angles of oscillation $\theta_0$ = $30^\circ$, $90^\circ$ and $180^\circ$.  }
\label{fig8}
\end{figure}

If one considers the apparent weight variation corresponding to oscillations of $60^\circ$, $180^\circ$or $360^\circ$ amplitude and with $\omega_{f}$ given by $\omega$ of Ec.\ref{eq:4}, one can visualize the fact that at the point of minimum potential energy, the oscillator has an apparent weight of its own that depends on the amplitude of the oscillation. Taking into account the amplitude limit, this apparent weight can become 2 times its own weight for a $60^\circ$ oscillation, 3 times for a $180^\circ$ oscillation, and 5 times for a full $360^\circ$ oscillation (Fig.\ref{fig8}).

\begin{figure}[htb!]
\centering
\includegraphics[width=75mm]{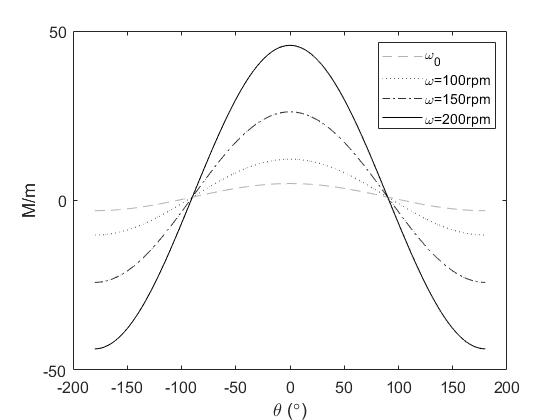}
\caption{Variation of the apparent weight on the gravity axis with respect to the initial angular velocity, $\unit[100]{rpm}$, $\unit[150]{rpm}$ and $\unit[200]{rpm}$.}
\label{fig9}
\end{figure}

\begin{figure}[htb!]
\centering
\includegraphics[width=75mm]{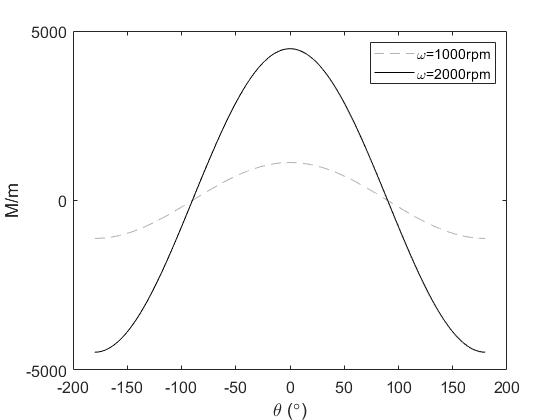}
\caption{Variation of the apparent weight on the gravity axis with respect to the initial angular velocity $\unit[1000]{rpm}$ and $\unit[2000]{rpm}$.}
\label{fig10}
\end{figure}

It can be visualized how the existing variation in the expression of the pendulum's apparent weight due to the introduction of energy evolves for different regimes of angular velocities $w_f$ from $\unit[100]{rpm}$ to $\unit[200]{rpm}$ (Fig.\ref{fig9}) and also for a higher velocity range such as $\unit[1000]{rpm}$ to $\unit[2000]{rpm}$ (Fig.\ref{fig10}).

The existence of a tendency in the high velocity range towards a harmonic behavior of the apparent weight can thus be verified. This is because, given its magnitude, the inertia component governs the motion, shielding the difference due to the oscillator's weight between the point of maximum and minimum potential, leading to an isotropic response throughout the oscillation.

\newpage

Table \ref{tab:1} shows the different angular velocities and energy increments, as well as the relative variation of the apparent weight between the points of maximum and minimum potential energy. The tendency towards such symmetry in the motion due to the increase of the angular velocities in the system can be clearly observed. 

\begin{table}[htb!]
\begin{center}
\begin{tabular}{| c | c | c | c | c | c |}
\hline
$\unit[\omega]{(rpm)}$ &  $\unit[\Delta E]{(J)}$ & M/m max & M/m min & $\Delta max-min(\%)$ \\ \hline
59,78 & 0 & 5  & -3  & 40 \\
100  & 35,23  & 12,19  & -10,19  & 16,40 \\
150 & 103,17 & 26,17 & -24,17 & 7,64 \\
200 & 199,72 & 45,76 & -43,76 & 4,37 \\
1000  & 5463,51 & 1120,00 & -1118,00  & 0,17 \\ 
2000  & 21912,85 & 4477.01 & -4475.01  & 0,04 \\ \hline
\end{tabular}
\caption{Relationship between the magnitude of the energy increment applied, the resulting angular velocity and the relative variation between  of maximum and minimum apparent weight points.}
\label{tab:1}
\end{center}
\end{table}

\subsubsection{Regimes of uniform acceleration}

The variation of the pendulum's apparent weight that occurs when it has a given constant acceleration will now be analyzed.. That is, the angular velocity will increase after each traveled degree.. In this case the angular velocity of the oscillator responds to Eq.\ref{eq:17} for an accelerated circular motion \cite{FundamentalsofPhysics,tipler2021fisica}.

\begin{equation}
    \omega_f^2=\omega_0^2+2\theta\alpha
    \label{eq:17}
\end{equation}

Where $\theta$ corresponds to the angular variation that the oscillator has traveled and $\alpha$ the angular acceleration with which it rotates. Taking into account the seesaw system (Fig.\ref{fig7}) the variation of the apparent weight along the oscillation (Ec.\ref{eq:18}) can be found through inserting Ec.\ref{eq:17} in Ec.\ref{eq:16}. 

\begin{equation}
{M(t)}=\frac{r_1}{r_2}(l\omega_0^2cos\theta+2l\theta\alpha cos\theta +g)m  
\label{eq:18}
\end{equation}

\begin{figure}[htb!] 
\centering
\includegraphics[width=75mm]{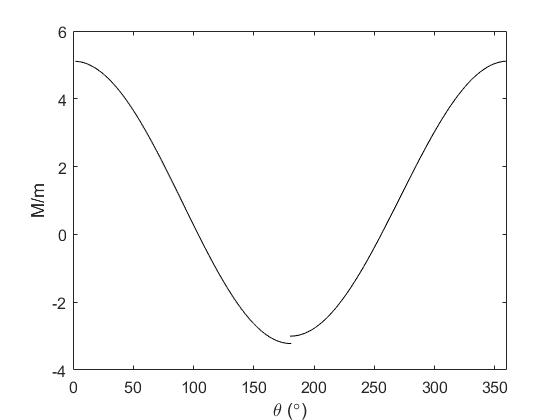}
\caption{Variation of the apparent weight of the pendulum over a complete oscillation of $360^\circ$ when the system has a constant angular acceleration $\alpha=\unit[1000]{rpm^2}$.}
\label{fig11}
\end{figure}

It is possible to visualize in Fig.\ref{fig11} the behavior of the apparent weight. An angular acceleration of $\unit[1000]{rpm^2}$ has been chosen to perform the calculation, with the aim of accentuating the representation of the features raised. Starting with $l=\unit[1]{m}$ and an initial angle of $180^\circ$ it can be observed that once the pendulum travels the $360^\circ$, the apparent weight at the end of the oscillation is $7\%$ higher than at the beginning, as visualized in the discontinuity of Fig.\ref{fig11}. From this relationship it follows that, after fulfilling an oscillation with an accelerated motion, the apparent weight of the pendulum will have changed with respect to the initial one even though the angular position coincides.\ The analyzed behavior will be amplified as the angular travel of the oscillator increases, i.e. during the successive cycles it completes, thus producing positive and negative modifications of the apparent weight with respect to its value at rest.

\break

\subsection{Geometry of dynamic configuration: accelerated motion by application of external force}

An accelerated motion of the oscillator by applying an external force on the system is now analyzed. With this aim, different scenarios, where distinct geometrical configurations of the point and direction of application of the said external force are evaluated, are going to be proposed.

\subsubsection{Angles and direction of application}

It is proposed to find the magnitude of the external force necessary for the oscillator to rotate with a certain value of angular acceleration starting from its natural angular velocity $\omega_{0}$. This force will depend on the direction in which it is applied on the oscillator, as well as on the angular instant in which it is applied. To illustrate the response of the system, two directions of application are chosen: normal and tangential with regard to the motion. This choice is explained by two factors: it simplifies the calculation and it shows the magnitude range in which the dynamic scheme is developed, since these directions correspond to the maximum and minimum projections of the inertia component respectively. In the first case, the magnitude of the external force necessary to accelerate the oscillator when this force is applied normal to the motion on the mass $M$, located at the other end of the seesaw,  is analyzed. The motion considered in the analysis has a uniformly accelerated character, so the angular velocity at each angular instant will correspond to the relation referred to in Ec.\ref{eq:17}. To visualize the magnitude of the force that will be necessary to apply as a function of the angular instant in which the oscillator is located, the condition of equilibrium of the seesaw is maintained, imposing that the torque $\vec{\tau}$ at both ends must be compensated. The value of $M$ along the oscillation is thus found for a motion without external acceleration (Eq.\ref{eq:16}), considering that $\theta_{0}=-180^\circ$ and a complete oscillation of $360^\circ$. Substituting the Ec.\ref{eq:16} in Eq.\ref{eq:19} it is found that, in the case where the external force is applied on $M$ in the direction of gravity, in order to obtain a motion with acceleration $alpha$, the variation of the apparent weight has to compensate the normal force (Eq.\ref{eq:20}).

Therefore, the thrust or normal force to be applied will have to be equal in magnitude to the desired apparent weight change (Eq.\ref{eq:20}), i.e. equalizing both forces and imposing the torque condition. 

\begin{equation}
r_{1}(ml\omega_f^{2}cos\theta(t)+mg)=r_{2}(Mg+F_{normal}) \\
\label{eq:19}
\end{equation}

\begin{equation}
\frac{\overrightarrow{F}_{normal}}{m}= \left (\frac{r_1}{r_2}(l\omega_0^2+2l\theta \alpha-4g) cos\theta \right ) \widehat{n}
\label{eq:20}
\end{equation}

In the second case, the magnitude of the force to be supplied if applied on the mass $m$ of the oscillator in the direction tangential to its motion is studied (Eq.\ref{eq:21}). The dynamic balance must include the contribution of the tangential projection of the oscillator's forces, considering that the angular acceleration is the ratio of the tangential acceleration to the pendulum length.

\begin{equation}
\frac{\vec{F}_{tangencial}}{m}=(g sen\theta +\alpha  l)\widehat{t}
\label{eq:21}
\end{equation}

\begin{figure}[htb!]
  \centering
    \includegraphics[width=80mm]{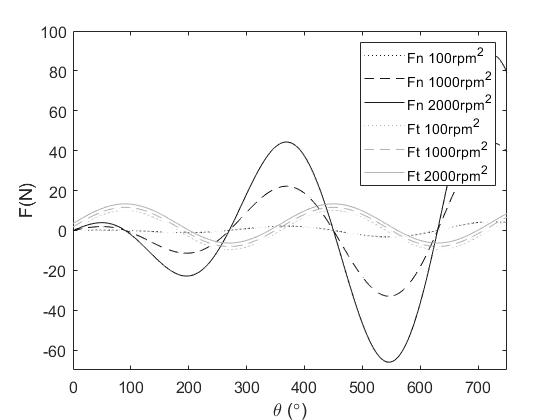}
    \caption{Application of the external force as a function of the angular instant required to generate the accelerated motion of the oscilator up to $\alpha=\unit[100]{rpm^2}$, $\unit[1000]{rpm^2}$ and $\unit[2000]{rpm^2}$ applied in the normal and tangential directions up to $750^\circ$.}
    \label{fig12}
    \end{figure}

\begin{figure}[htb!]
  \centering
    \includegraphics[width=80mm]{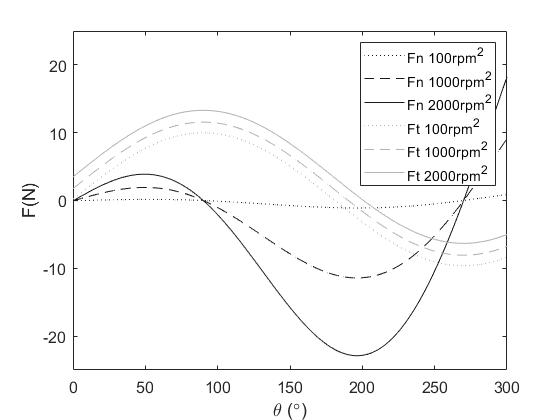}
    \caption{Application of the external force as a function of the angular instant required to generate the accelerated motion of the oscilator up to $\alpha=\unit[100]{rpm^2}$, $\unit[1000]{rpm^2}$ and $\unit[2000]{rpm^2}$ applied in the normal and tangential directions up to $300^\circ$.}
    \label{fig13}
\end{figure}

The results of this analysis for different values of angular acceleration are displayed in Fig.\ref{fig12} and Fig.\ref{fig13}. Specifically, the force that needs to be applied in the tangential and normal directions to the oscillator's motion in order to obtain angular accelerations of 100, 1000 and $\unit[2000]{rpm^2}$ is shown.

The main feature observed in these relations Eq.\ref{eq:20} and Eq.\ref{eq:21}, which can be visualized in Fig.\ref{fig12} and Fig.\ref{fig13}, is that the external force modifies its sign synchronously with the period of the oscillator.
That is to say, in order to achieve an accelerated motion in the oscillator it will be necessary to perform on it a pushing or braking force depending on the angular interval in which it is applied. The fundamental implication of this result is the interpretation of that negative component of the apparent force or weight. There are geometries for which, depending on the angular velocity, the angular instant and the magnitude of the load, the application of the latter on the mass $M$ accelerates the oscillator instead of slowing it down. That is to say, the meaning of the load varies depending on the geometry and the instant in which the motion of the system is found. 

This leads to scenarios where, under the adequate conditions, not only there are not any losses due to energy applications but also it is possible to extract energy from the system. The difference in the application of the thrust force, whether tangential or normal, highlights the importance of the direction of application as well as the angular instant at which such force is applied. Therefore a normal thrust force will always be greater in magnitude than a tangential one due to the inertial weight component of the mass $m$ of the oscillator. Observing Eq.\ref{eq:20} and Eq.\ref{eq:21} it becomes clear that, for the tangential thrust force, the projection of the rotational inertia component is zero. That is, the resultant will not reflect this change because the system will not increase its angular velocity along its path. These results can be observed in Fig.\ref{fig13}, where for a tangential thrust the amplitude does not vary along its path. On the contrary, for a normal thrust its amplitude will be increased along its path. These changes are more significant for high angular accelerations, since they entail larger magnitude increases of the system's angular velocity.   

\subsubsection{Resultant acceleration vector}

So far it has been considered in the analysis that the resultant angular acceleration has the direction of the spin axis of the oscillator above $M$, but other conceptual scenarios, in which this resultant is forced to take another direction can also be opened, evaluating the magnitude of external force that will be necessary in order to cause the acceleration of the system above $m$. Two possible scenarios in which the acceleration is fixed in a different direction will now be analyzed. 

In the first case, for the sake of simplicity, it is proposed that the resultant is restricted to the axis of gravity, in order to evaluate the maximum contribution that the motion may have in that direction. The apparent weight variation that will be necessary to contribute along the oscillation is calculated (Eq.\ref{eq:22}).

\begin{figure}[htb!] 
\centering
\includegraphics[width=80mm]{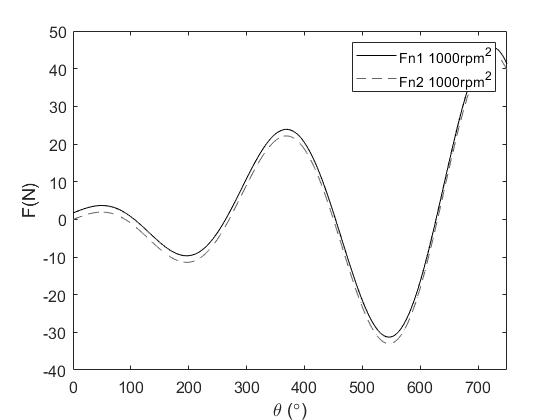}
\caption{Comparison of the magnitude required to produce the resultant acceleration $\alpha=\unit[1000]{rpm^2}$ in the axis of rotation described in Fig.\ref{fig5} and in the case where it is restricted to the gravity or normal axis. }
\label{fig14}
\end{figure}

It can be visualized in Fig.\ref{fig14} how forcing a resultant behavior in the normal direction with regard to mass $m$ makes a negligible difference compared to the normal direction with regard to mass $M$. In particular, for the case where the length of the pendulum is $l=\unit[1]{m}$, the translation with respect to the acceleration in the direction of the spin axis is less than $2N$.

\begin{equation}
\frac{\overrightarrow{F}_{normal}}{m}=\left (\frac{r_1}{r_2}((l\omega_f^2 - 4g)cos\theta+\alpha)\right)\widehat{n}
\label{eq:22}
\end{equation}

In the second case the direction of the resultant acceleration is set in a tangential direction with regard to the pendulum's motion (Eq.\ref{eq:23}).

\begin{equation}
\frac{\vec{F}_{tangencial}}{m} = (g sen\theta + \alpha)\widehat{t}
\label{eq:23}
\end{equation}

It is implied that the difference between the normal and tangential resultant will be proportional to the pendulum's length. In this paper it has been assumed that the pendulum's length $l=\unit[1]{m}$ in order to make the conceptual implications of the calculations clearer. For this case, the behaviors in both projections coincide in magnitude.

These two assumptions are particular cases of the infinite directions that can be theoretically posed. It is thus shown that the possibility of extracting a force from the system, when it is accelerated by the application of a charge under certain geometrical configurations, is available for all the possible directions the resultant force vector can adopt.
\section{Conclusions}

Since it does not have a rectilinear motion, the system undergoes a centrifugal force due to its rotational inertia in the direction of the rod and towards the outside of the axis of rotation. On the other hand, the velocity of the mass $m$ is perpendicular to the inertia component along the oscillation, i.e. tangential to the motion.

Considering these directions the necessary forces, both normal and tangential, to be applied in order to produce a given angular increase in the amplitude of the oscillation have been analyzed. A feature in the present investigation is the difference in apparent weight corresponding to the amplitude of the oscillator, which makes it possible to observe that for higher amplitude ranges, a greater force must be applied to move the center of mass. This variation of apparent weight can be up to 5 times greater for a complete oscillation of $360^\circ$.

The accelerated motion of the pendulum can be achieved through the direct introduction of energy into the system or applying some force on it. The magnitude of force is reflected in the different geometries of application, depending on direction sense and point of application. Therefore, the effect of a normal thrust will be much greater in magnitude than the effect of a tangential thrust due to the inertia component of the oscillator mass itself.

It is demonstrated in the present work that the magnitude of the external force necessary to cause the acceleration of the oscillator takes negative values for certain angular regions. That is to say, an external load or brake applied synchronously to the period of the pendulum may result, under the right conditions, in an increase of the angular acceleration of the motion and, therefore, in a possible extraction of energy from the system. 

It is thus exposed the behavior in which certain configurations can be extracted from the system, such as the ability to exert a force resulting in an increase in the acceleration of the oscillator, and that is extensible to all geometric results that may occur.

\section{Acknowledgments}

We would like to thank our colleagues of Zoocánica S.Coop. and the cooperative entities that have set it up, because with their work they make possible this and other scientific and technological research projects carried out in a social economy framework, betting on a transfer model that seeks the public interest and the transformation of the productive framework.

%\section*{References}

\bibliography{mybibfile}

\begin{thebibliography}{10}
\expandafter\ifx\csname url\endcsname\relax
  \def\url#1{\texttt{#1}}\fi
\expandafter\ifx\csname urlprefix\endcsname\relax\def\urlprefix{URL }\fi
\expandafter\ifx\csname href\endcsname\relax
  \def\href#1#2{#2} \def\path#1{#1}\fi

\bibitem{FundamentalsofPhysics}
J.~W. D.~Halliday, R.~Resnick, Fundamentals of Physics, Wiley, 2007, 10th
  edicion.

\bibitem{cespedesprimer}
F.~D. C{\'e}spedes, El primer experimento de galileo galilei., Revista
  Latinoamericana de Ensayo Fundada en Santiago de Chile.

\bibitem{galileo}
F.~A. Buyse, Galileo galilei, holland and the pendulum clock, O que nos faz
  pensar 26~(41).

\bibitem{modinoemmy}
S.~M. Modino, Emmy noether y su impacto en la f{\i}sica te{\'o}rica.

\bibitem{libropenduloresumen}
G.~L. Baker, J.~A. Blackburn, The pendulum a case study in physics, Oxford
  University Press, 2005.

\bibitem{tipler2021fisica}
P.~A. Tipler, G.~Mosca, F{\'\i}sica para la ciencia y la tecnolog{\'\i}a.
  Volumen 1A: Mec{\'a}nica, Vol.~1, Reverte, 2021.

\bibitem{empuje}
P.~Binder, J.~C. Bragg, The elastic ballistic pendulum, Physics Education
  54~(5) (2019) 053003.

\bibitem{columpio}
W.~B. Case, The pumping of a swing from the standing position, American Journal
  of Physics 64~(3) (1996) 215--220.

\bibitem{columpio2}
C.~P. y~Julio~Gratton,
  \href{https://anales.fisica.org.ar/journal/index.php/analesafa/article/view/264}{El
  columpio}, ANALES AFA 16~(1).
\newline\urlprefix\url{https://anales.fisica.org.ar/journal/index.php/analesafa/article/view/264}

\bibitem{columpio3}
S.~Wirkus, R.~Rand, A.~Ruina, How to pump a swing, The College Mathematics
  Journal 29~(4) (1998) 266--275.

\bibitem{Tea1968PumpingOA}
P.~Tea, H.~Falk, Pumping on a swing, American Journal of Physics 36 (1968)
  1165--1166.

\end{thebibliography}

\end{document}